\documentclass[aps,showpacs,amsmath,amssymb]{revtex4}
\usepackage{graphicx}
\usepackage{dcolumn}
\usepackage{bm}
\usepackage{xcolor}
\usepackage{epsfig}
\usepackage{multirow}
\usepackage{booktabs}

\begin{document}
\title{
Quasi-normal modes and absorption section from regular black holes immersed in perfect fluid dark matter
}

\author{L. O. T\'ellez Tovar$^1$}
\email{osvaldo_tellez@uaeh.edu.mx}
\author{Omar Pedraza$^1$}
\email{omarp@uaeh.edu.mx}
\author{L. A. L\'opez$^1$}
\email{lalopez@uaeh.edu.mx}
\author{R. Arceo$^2$}
\email{roberto.arceo@unach.mx}

\affiliation{$^1$ \'Area Acad\'emica de Matem\'aticas y F\'isica, UAEH, 
Carretera Pachuca-Tulancingo Km. 4.5, C. P. 42184, Mineral de la Reforma, Hidalgo, M\'exico.}

\affiliation{$^{2}$ Facultad de Ciencias en F\'isica y Matem\'aticas, Universidad Aut\'onoma de Chiapas, C. P. 29050, Tuxtla Guti\'errez, Chiapas, M\'exico.}


\begin{abstract}

We study scalar, electromagnetic and gravitational perturbations of three families of regular black holes, Hayward, Bardeen, and Ayon-Beato-García immersed in a Perfect Fluid Dark Matter (PFDM) background. Using both the sixth-order WKB approximation and the Asymptotic Iteration Method (AIM), we determine the corresponding quasi-normal modes  and absorption cross sections. By fixing the horizon radius, we analyze the dependence of the parameters  of the black holes and the parameter of PFDM. The computed quasi-normal modes confirm the dynamical stability of the black holes under scalar, electromagnetic and gravitational perturbations. Furthermore, both the real and imaginary parts of the quasi-normal modes increase with the multipole number $l$, corresponding to higher oscillation frequencies and faster damping rates. On the other hand, absorption cross sections are compared under similar parametric conditions. The results demonstrate that PFDM  influences the dynamical properties of regular black holes. 
\end{abstract}

\maketitle
\section{Introduction}

Estimates indicate that the universe is composed of approximately $ 68 \% $ dark energy, $ 27 \%$ dark matter, and only $5 \%$ baryonic (ordinary) matter \cite{Planck:2018vyg} \cite{Planck:2019nip}. As black holes are recognized as fundamental constituents of the universe, the study of their interaction and coexistence with dark matter and dark energy has become an area of considerable interest in theoretical and observational physics.

A notable contribution in this direction is the model proposed by Kiselev \cite{Kiselev:2002dx}, which describes black hole solutions immersed in dark energy. Within this framework, various types of black holes have been extensively analyzed \cite{Ghaderi:2017wvl} \cite{Pedraza:2020uuy} \cite{Saleh:2018hba}, leading to a deeper understanding of the influence of dark energy on black holes geometry and dynamics.

On the other hand, the earliest evidence supporting the existence of dark matter emerged from the observed flattening of rotation curves in spiral galaxies. Expanding on this, Kiselev \cite{Kiselev:2003ah} also introduced a model that considers a spherically symmetric black hole solution immersed in a Perfect Fluid Dark Matter (PFDM) environment and studied in \cite{Li:2012zx}. This approach has been employed to investigate a range of BHs configurations, for example  \cite{Zhang:2020mxi} \cite{Abbas:2023pug} \cite{Anjum:2023axh}.  On the other hand, also models with black holes immersed in dark matter have been studied, for example solutions with dark matter term from pure Weyl gravity  without the Einstein term \cite{Konoplya:2025mvj} \cite{Konoplya:2022hbl}.

As is well known, the study of quasi-normal modes provides valuable insights into the properties of black holes in various scenarios, as these modes are closely related to the parameters that describe the black hole. Additionally, analysing quasi-normal modes allows for the determination of oscillation frequencies through the real part, and the assessment of black holes stability via the imaginary part. The study of quasi-normal modes in black holes surrounded by dark energy or dark matter has been addressed in several works, such as \cite{Pedraza:2021hzw} \cite{Jusufi:2019ltj} \cite{Xu:2017bpz}.

Also the analysis of the Schrödinger-type differential equation, which emerges when considering radial perturbations in the black hole space-time, plays a fundamental role beyond the determination of quasi-normal modes. The same formalism can be extended to compute scattering and absorption cross sections. The quasi-normal modes, scattering and absorption cross sections are for gaining deeper insights into black holes properties. For example, studies of how dark energy affects the scattering and absorption sections of black holes have been studied in \cite{Lopez:2021ujg} \cite{Ramirez:2021ibk}.

For the reasons mentioned above, the purpose of this contribution is to investigate the effects of dark matter on black holes, by means of quasi-normal modes and absorption sections, especially considering the Hayward, Bardeen and Ayon-Beato-Garcia solutions. The absorption and scattering cross sections for Hayward were calculated in \cite{dePaula:2023muc}, absorption cross section of Bardeen surrounded by perfect fluid dark matter in \cite{Rincon:2025buq} and the scalar absorption in \cite{Paula:2020yfr}.

This paper is organized as follows. In Section \ref{sec.hori} we give a brief description of the Hayward, Bardeen and Ayon-Beato-Garcia black holes  immersed in PFDM and we analyzed the event horizons. Specifically, we study the parameter region where the black holes have two or one horizons. In Sec. \ref{sec.ep} we describe the scalar, electromagnetic and gravitational perturbations of a black hole. In Sec. \ref{sec.qnm} we present the results for quasi-normal modes using two methods: WKB approach and AIM. Section \ref{acs} is devoted to the calculations of the adsorption section of the Hayward, Bardeen and ABG  immersed in PFDM. Finally, we present our conclusions in Sec. \ref{conclu}.   
    
\section{Black holes in perfect fluid dark matter}\label{sec.hori}

Let us consider the space-time line element of spherically symmetric black hole immersed in PFDM can be expressed in the form:
\begin{equation}\label{mfa}
	ds^2=-f(r)dt^2+\frac{dr^2}{f(r)}+r^2d\theta^2+r^2\sin^2\theta d\phi^2\,,
\end{equation}
where the metric function is given by 
\begin{equation}\label{ec.rfc}
	f(r)=1-\frac{2m_i(r)}{r}\,,
\end{equation}
where the index $i=1,2,3$ corresponds to the Hayward, Bardeen, and ABG, respectively.

For the Hayward  \cite{Hayward:2005gi}, we have
\begin{equation}\label{key}
m_1(r)=\frac{M_1r^3}{r^3+2M_1q_1^2}-\frac{\alpha}{2}\ln\left(\frac{r}{\left|\alpha\right|}\right)	\,,
\end{equation} 
for the Bardeen \cite{Bardeen1}, we can express $m_2(r)$ as
\begin{equation}\label{key}
	m_2(r)=\frac{M_2r^3}{\left(r^2+q^2_2\right)^{3/2}}-\frac{\alpha}{2}\ln\left(\frac{r}{\left|\alpha\right|}\right)	\,,
\end{equation} 
and for the ABG  \cite{Ayon-Beato:1998hmi}, the function $m_3(r)$ can be written as:
\begin{equation}\label{key}
	m_3(r)=\frac{M_3r^3}{\left(r^2+q_3^2\right)^{3/2}}
	-\frac{q_3^2r^3}{2\left(r^2+q_3^2\right)^2}
	-\frac{\alpha}{2}\ln\left(\frac{r}{\left|\alpha\right|}\right)\,.
\end{equation} 

Here, $M_{i}$ ($i=1,2,3$) are the masses of the BHs, $q_1$ is a parameter related to the cosmological constant, $q_2$ is the magnetic charge, $q_3$ is the electric charge and the parameter $\alpha$ is associated with the intensity of the perfect fluid dark matter.
The constant $\alpha$ is related to the components of the energy-momentum tensor of PFDM $T^{\mu}_{\nu}=diag(- \rho,P_{r},P_{\theta},P_{\phi})$ \cite{Zhang:2020mxi} \cite{Das:2021otl} as:

\begin{equation}
-\rho=P_{r}=\alpha / 8 \pi r^{3}  \;\;\;\;\;\ P_{\theta}=P_{\phi}=-\alpha/ 8 \pi r^{3} \,,
\end{equation}
where $\rho$ is the density, $P_{r}$ the radial pressure, $P_{\theta}$ and $P_{\phi}$ the tangential pressures. Therefore, we shall restrict ourselves to the case $\alpha<0$ which gives positive energy density.

The horizons of the black holes surrounded by PFDM are determined by the positive roots of the equation $f(r_h) = 0$, then the event horizons are the roots of
\begin{eqnarray}
\left[r_h+\alpha\ln\left(\frac {r_h}{\left| \alpha\right|}\right)\right]\left(2M_1q_1^2+r_h^3\right)-2M_1 r_h^3&=&0\,,\qquad\text{Hayward}\,,
\label{ec.heH}	\\
\left[r_{h}+\alpha\ln\left(\frac {r_h}{\left| \alpha\right|}\right)\right] 
\left( q_2^2+r_h^2\right) ^{\frac{3}{2}}-2M_2r_h^3&=&0\,,\qquad\text{Bardeen}\,
\label{ec.heB}\,,\\
\left[r_{h}+\alpha\ln\left(\frac {r_h}{\left| \alpha\right|}\right)\right]\left(r^2_{h}+q_3^2\right)^2-2M_3r^3_{h}\sqrt{r^2_{h}+q_3^2}+q_3^2r^2_{h}&=&0\,,\qquad\text{ABG}\,.
\end{eqnarray}
In an appropriate parametric region, the BHs immersed in perfect fluid dark matter can have two horizons: the inner horizon $r_{-}$ and the event horizon $r_+(r_-\leq r_+)$. Clearly, the number of horizons depends on the choice of the parameter values $M_{i}$, $q_{i}$ and $\alpha$. 

\begin{figure}[!h]
	\centering
	\includegraphics[scale=0.92]{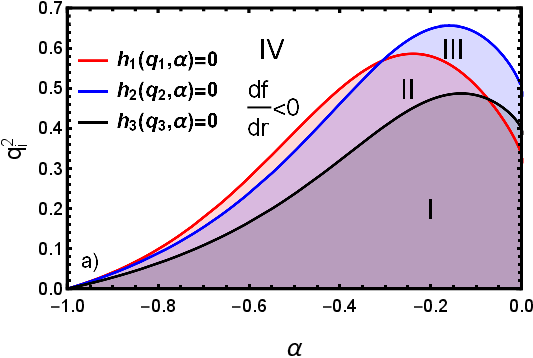}
	\includegraphics[scale=0.94]{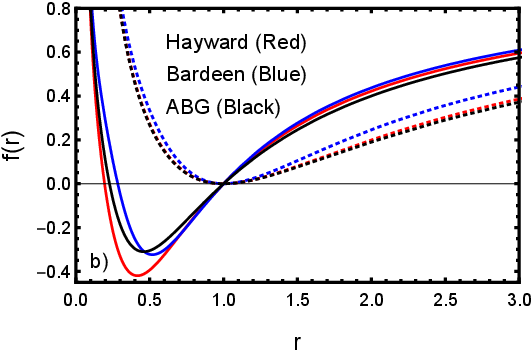}
	\caption{ a) The parametric region that allows the event horizon $r_+$ and inner horizon $r_-$. The region below the curves $h_1$ (for values ($\alpha,q_1^2$)), $h_2$ (for values ($\alpha,q_2^2$)) and $h_3$ (for values ($\alpha,q_3^2$)) have two event horizons, while for all values ($\alpha,q_1^2$), ($\alpha,q_2^2$), ($\alpha,q_3^2$) where $h_1=0$, $h_2=0$ and $h_3=0$, the BHs have only one horizon $r_+$.  b) The behaviour of the metric function for Hayward, Bardeen and ABG black holes. The solid lines are $\alpha=-0.1$ and $q_1^2=q_2^2=q_3^2 =0.1$, while the dotted lines are for $\alpha=-0.3$, $q_1^2=0.571789$, $q_2^2=0.575457$ and $q_3^2=0.406527$.}
	\label{f1}
\end{figure}

To determine the range of values for $M_{i}$, $q_{i}$ and $\alpha$, we make use of the method described in \cite{Liu:2021fzr} \cite{Liu:2020evp}. Thus, we consider $f(r_h)=0$, where for convenience, we fix the event horizon as $r_+=1$ in this work. Then, from (\ref{ec.heH}), the masses of BHs can be expressed as 
\begin{eqnarray}
M_1(q_1,\alpha)&=&
\frac{1+\alpha\ln\left(\frac{1}{\left|\alpha\right| }\right)}{2-2q_1^2\left[1+\alpha\ln
	\left(\frac{1}{\left|\alpha\right| }\right)\right]}\,,\qquad\text{Hayward}\,,
\label{ec.maH}\\
M_2(q_2,\alpha)&=&\frac{\left(q_2^2+1\right)^{3/2}}{2} \left[1+\alpha  \ln
\left(\frac{1}{| \alpha | }\right)\right]
\,,\qquad\text{Bardeen}\,,\label{ec.maB}\\
M_3(q_3,\alpha)&=&
\frac{\left[1+\alpha  \ln
	\left(\frac{1}{| \alpha | }\right)\right]\left(q_3^2+1\right)^2+q_3^2}{2\sqrt{q^2_3+1}}
\,,\qquad\text{ABG}\label{ec.maHABG}\,.
\end{eqnarray}
To continue with the analysis, it is necessary to study the parametric region of values $(\alpha,q_1^2)$ for Hayward, $(\alpha,q_2^2)$ for Bardeen and $(\alpha,q_3^2)$ for ABG. Thus we introduce $M_1(q_1,\alpha)$, $M_2(q_2,\alpha)$ and $M_3(q_3,\alpha)$ in $f(r)$. Then, the parametric region in which the black hole event horizons are allowed can be determined by requiring $\frac{d}{dr}f(r_+)>0$ (the black hole temperature is positive). This leads to

\begin{eqnarray}\label{ec.M}
h_1(q_1,\alpha)=1+\alpha-3q_1^2\left[1+\alpha  \ln
\left(\frac{1}{| \alpha | }\right)\right]^2&>&0
\,,\qquad\text{Hayward}\,,\\
h_2(q_2,\alpha)=q_2^2\left[
\alpha-2+3\alpha\ln \left(| \alpha |
\right)\right]+1+\alpha&>&0
\,,\qquad\text{Bardeen }\,,\\
h_3(q_3,\alpha)=\alpha+3\alpha(q_3^2+1)^2 q_3^2 \ln(|\alpha|)+(\alpha-2)q_3^6+3 (\alpha -1)
q_3^4+3 \alpha q_3^2-q_3^2+1&>&0\,,\qquad\text{ABG}\,.
\end{eqnarray}  
The behavior of $q_i^2$ ($i=1,2,3$) as function of $\alpha$ is shown in Fig. \ref{f1} a). In region, I, the ABG BH has two horizons, while for values of ($\alpha,q^2_3$) where $h_3(\alpha,q_3)=0$, the ABG BH has one horizon (see Fig. \ref{f1} b)). This behavior is similar for Hayward and Bardeen BH, i.e. in region II the Bardeen BH and in region III the Hayward BH are two horizons.      

\section{Massless scalar, electromagnetic and gravitational perturbations}\label{sec.ep}

The analysis of quasi-normal modes  frequencies in black holes serves as a powerful tool for probing both the intrinsic properties of the black hole and the nature of its possible emissions. When the black hole exhibits radial symmetry, the eigenvalue problem associated with quasi-normal modes takes a form analogous to the Schr\"odinger equation. In this section, we provide a brief overview of the behavior of scalar, electromagnetic and gravitational perturbations around the BHs immersed in PFDM. 

The Klein-Gordon equation for a massless scalar field in curved space-time is given by
\begin{equation}\label{ec.kg}
\frac{1}{\sqrt{-g}}\partial_{\mu}\left(\sqrt{-g}\,g^{\mu\nu}\partial_{\nu}\psi\right)=0\,.
\end{equation}
By using the ansatz for the scalar field $\psi$
\begin{equation}\label{ec.ant}
\psi=e^{-i\omega t}Y_{lm}(\theta,\phi)\frac{R(r)}{r}\,.
\end{equation}
Introducing (\ref{ec.ant}) in (\ref{ec.kg}), we obtain the following Schr\"odinger-type equation
\begin{equation}\label{ec.ts}
\frac{d^2R(r_*)}{dr_*^2}+\left[\omega^2-V_{\text{s}}(r)\right]R(r_*)=0\,,
\end{equation}
where $\omega = \omega_r + i\omega_i$, with $\omega_r$ representing the oscillation frequency of the black hole, which is always positive, while $\omega_i$ is associated with the stability of the black hole. The effective potential $V_{\text{s}}$ is given by,
\begin{equation}\label{ec.ps}
V_{\text{s}}(r)=f(r)\left[\frac{l(l+1)}{r^2}+\frac{1}{r}\frac{df(r)}{dr}\right]\,,
\end{equation} 
here $r_*$ is the tortoise coordinate given by
\begin{equation}
dr_*=\frac{dr}{f(r)}\,,
\end{equation}
and $l$ is the spherical index. 

On the other hand, the electromagnetic field in curved space-time follows the next expression
\begin{equation}\label{ec.ef}
\frac{1}{\sqrt{-g}}\partial_{\mu}\left(\sqrt{-g}\,F_{\nu\lambda}g^{\nu\alpha}g^{\lambda\mu}\right)=0\,,
\end{equation}  
with $F_{\nu\lambda}=\partial_{\nu}A_{\lambda}-\partial_{\lambda}A_{\nu}$ and $A_{\lambda}$ is the four vector potential.  After the separation of variables, the radial part of the Eq. (\ref{ec.ef}) can express the Schr\"odinger-type equation (\ref{ec.ts}) again, but with a different potential given by
\begin{equation}\label{ec.pem}
V_{\text{em}}(r)=f(r)\frac{l(l+1)}{r^2}\,.
\end{equation}

The generalized form of the effective potential \cite{Nomura:2005dn} for different test fields can be written as

\begin{eqnarray}\label{Vge}
	V(r)&=&f(r)\left[\frac{l(l+1)}{r^2}+\left(1-a^2\right)\frac{2m_{i}(r)}{r^3}+\left(1-a\right)\left(\frac{1}{r}\frac{df(r)}{dr}-\frac{2m_{i}(r)}{r^3}\right)
	\right]
	\,.
\end{eqnarray}

Here, $a=0,1,2$  denotes the spin of the perturbation, corresponding to scalar, electromagnetic, and gravitational fields, respectively. From Eq. (\ref{Vge}), we observe that the effective potential $V(r)$ depends on the parameters $q_{i}$, $\alpha$, $M_{i}$, and the angular harmonic index $l$ (with $l \geq a$).

\begin{figure}[!h]
	\centering
	\includegraphics[scale=0.93]{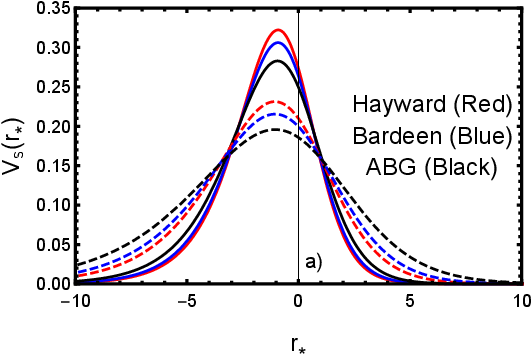}
	\includegraphics[scale=0.93]{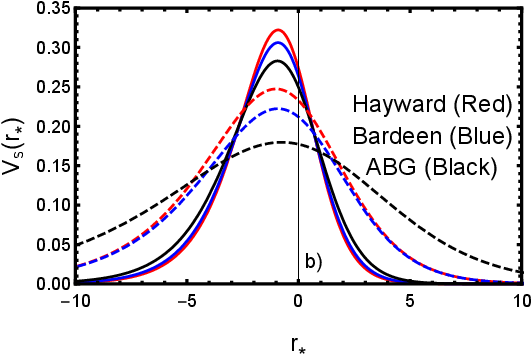}
	\caption{
		a) The behavior of $V_s(r_*)$ with $r$ for scalar perturbations, where the solid line corresponds to $\alpha=-0.1$ and the dot-dashed line corresponds to $\alpha=-0.4$ Here $q_1^2=q_2^2=q_3^2=0.1$ and $l=1$. b) The behavior of the effective potential $V_s(r_*)$ is shown for different values of $q_1^2$, $q_2^2$ and $q_3^2$, with $\alpha=-0.1$ and $l=1$. The solid lines are for $q_1^2=q_2^2=q_3^2=0.1$, while the dot-dashed lines are for $q_1^2=q_2^2=q_3^2=0.3$. }
	\label{f2}
\end{figure}

\begin{figure}[!h]
	\centering
	\includegraphics[scale=0.93]{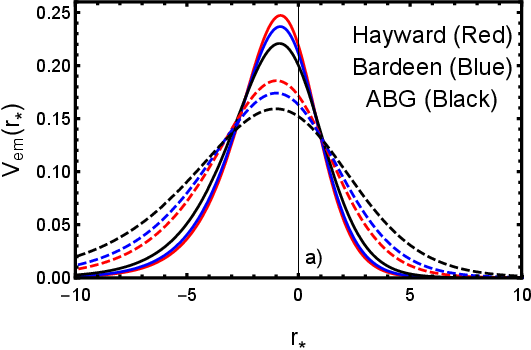}
	\includegraphics[scale=0.93]{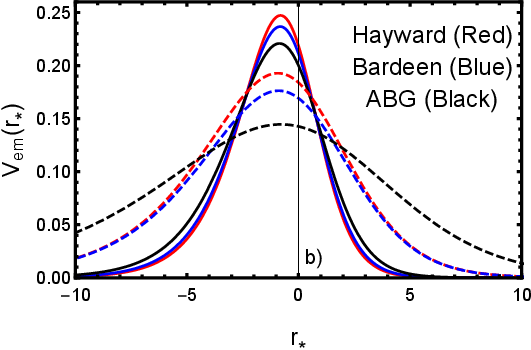}
	\caption{
		a) The behavior of $V_{em}(r_*)$ with $r$ for electromagnetic perturbations, where the solid line corresponds to $\alpha=-0.1$ and the dot-dashed line corresponds to $\alpha=-0.4$ Here $q_1^2=q_2^2=q_3^2=0.1$ and $l=1$. b) The behavior of the effective potential $V_{em}(r_*)$ is shown for different values of $q_1^2$, $q_2^2$ and $q_3^2$, with $\alpha=-0.1$ and $l=1$. The solid lines are for $q_1^2=q_2^2=q_3^2=0.1$, while the dot-dashed lines are for $q_1^2=q_2^2=q_3^2=0.3$.}
	\label{f3}
\end{figure}

\begin{figure}[!h]
	\centering
	\includegraphics[scale=0.93]{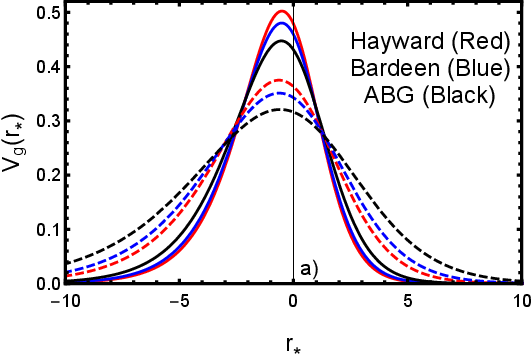}
	\includegraphics[scale=0.93]{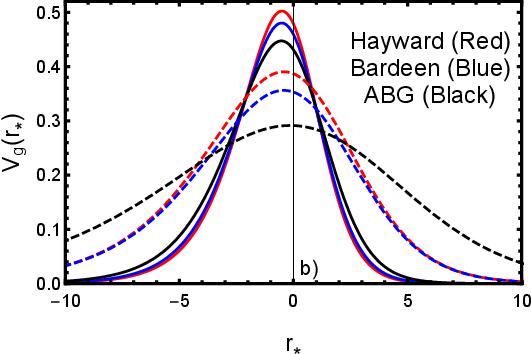}
	\caption{a) The behavior of $V_{g}(r_*)$ with $r$ for gravitational perturbations, where the solid line corresponds to $\alpha=-0.1$ and the dot-dashed line corresponds to $\alpha=-0.4$ Here $q_1^2=q_2^2=q_3^2=0.1$ and $l=2$. b) The behavior of the effective potential $V_{g}(r_*)$ is shown for different values of $q_1^2$, $q_2^2$ and $q_3^2$, with $\alpha=-0.1$ and $l=2$. The solid lines are for $q_1^2=q_2^2=q_3^2=0.1$, while the dot-dashed lines are for $q_1^2=q_2^2=q_3^2=0.3$.}
	\label{f3.1}
\end{figure}

We plot the effective potentials $V_{\text{s}}(r_*)$, $V_{em}(r_*)$  and $V_{g}(r_*)$ in Figs. \ref{f2}, \ref{f3} and $\ref{f3.1}$ respectively. We notice that when the factor $\alpha$ decreases, the peak of the potential $V_s(r_*)$ also decreases (see Fig. \ref{f2} a)). While when $q_{i}^2$ increase, the maximum height of the potential $V_s(r_*)$ decreases (see Fig. \ref{f2} a)). The effective potential of the electromagnetic and gravitational perturbations also have a behavior similar to $V_{\text{s}}(r_*)$, as can be seen in Fig. \ref{f3} and $\ref{f3.1}$.


However, the maximum height of the potential is higher for the Hayward black hole and lower for the ABG black hole for all three types of perturbations, regardless of the values of the parameters $q_{i}$ and $\alpha$ (see Figures \ref{f2}-\ref{f3.1}).

\section{Quasi-normal modes}\label{sec.qnm}

Now the Eq. (\ref{ec.ts}) can be solved to obtain the quasi-normal modes by imposing that the wave at the event horizon is purely incoming and the wave at the spatial infinity is purely outgoing. Mathematically, we can write the above-mentioned boundary conditions as, 
\begin{equation}\label{key}
R=\begin{cases}
e^{-i\omega r_*}\sim \left(r-r_+\right)^{-i\omega/2\kappa_+}&\text{if}\quad r_*\to-\infty (r\to r_+)\\	
e^{i\omega r_*}\sim e^{i\omega\int^r \frac{dx}{f(x)}}&\text{if}\quad r_*\to+\infty(r\to \infty)\\
\end{cases}\,,
\end{equation}
where $\kappa_+$ is the surface gravity on the event horizon defined as
\begin{equation}\label{ec.gs}
\kappa_+=\left.\frac{1}{2}\frac{df(r)}{dr}\right|_{r=r_+}\,.	
\end{equation}

Several numerical and semi-analytical techniques have been developed to compute the frequencies of quasi-normal modes. Among the most commonly employed is the sixth-order WKB approximation method formulated by Konoplya \cite{Konoplya:2003ii} \cite{Konoplya:2019hlu}. 

The formula for the quasi-normal modes is
\begin{equation}
\frac{i\left(\omega^2-V_0\right)}{\sqrt{-2V''_0}}-\sum_{j=2}^{6}\Lambda_j=n+\frac{1}{2}\,,	
\end{equation}
where $n$ is the overtone number, $V_0$ and $V''_0$ are the maximum potential and the second derivative of the potential where the maximum occurs. The expressions $\Lambda_j$ are given in  \cite{Konoplya:2003ii} (for $j=4,5,6$). However, the WKB approach becomes unreliable for modes where the overtone number $n$ exceeds the angular momentum quantum number $l$. Alternatively, a more recently developed method, known as the Asymptotic Iteration Method, has shown improved efficiency.  It is worth mentioning that the literature includes other methods for calculating quasi-normal modes. A brief summary of some of these methods can be found in \cite{Konoplya:2011qq}.

\subsection{The asymptotic iteration method (AIM)}

In this subsection, the application of the AIM method to obtain the oscillation frequencies of the quasi-normal modes of BH is presented in a summarized manner.

To carry out the AIM approach is convenient to implement the change of variable $r=1/\xi$. In the new variable, the Eq. (\ref{ec.ts}) now can be expressed as,
\begin{equation}\label{ec.aim1}
\frac{d^2R}{d\xi^2}+\frac{p'}{p}\frac{dR}{d\xi}+\left[
\omega^2-v\left(\xi\right)
\right]	R=0\,,
\end{equation}
where 

\begin{eqnarray}
p\left(\xi\right)&=&\xi^2-2\xi^3 m\left(\xi\right)\,,\label{ec.dp}\\
v\left(\xi\right)&=&\frac{l(l+1)}{p}
+
\left(1-a^2\right)\frac{2m(\xi)\xi}{p}
-
\left(1-a\right)\left[\frac{p'}{p\xi}-\frac{2}{\xi^2}+\frac{2m(\xi)\xi}{p}\right]
\,.
\end{eqnarray}

Here, $a=0$ represents the scalar field, $a=1$ refers to the electromagnetic field and $a=2$ the gravitational field. To apply the boundary conditions, it is clear that when $r\to r_+$, $\psi$ behaves like $e^{-i\omega r_*}\sim \left(r-r_+\right)^{-i\omega/2\kappa_+}\sim\left(\xi-\xi_+\right)^{-i\omega/\kappa_+}$ and $\psi\sim e^{i\omega r_*}\sim e^{i\omega\int\frac{dr}{f(r)}}\sim e^{-i\omega\int\frac{d\xi}{p(\xi)}}$ at $r\to\infty$. 

To solve Eq. (\ref{ec.aim1}) with AIM, we propose the following relation
\begin{equation}\label{ec.rfr}
R(\xi)=\left(\xi-\xi_+\right)^{-i\omega/\kappa_+}e^{-i\omega\int\frac{d\xi}{p(\xi)}}\chi(\xi)\,.
\end{equation}
Introducing (\ref{ec.rfr}) into Eq. (\ref{ec.aim1}), we have the following expression
\begin{equation}\label{ec.aim}
\frac{d^2\chi(\xi)}{d\xi^2}=\lambda_0(\xi)\chi(\xi)+s_0(\xi)\chi(\xi)\,,
\end{equation}
where
\begin{eqnarray}
\lambda_0\left(\xi\right)&=&
\frac{2i\omega}{\kappa_+\left(\xi-\xi_+\right)}-\frac{p'-2i\omega}{p}
\label{l0} \,,\\
s_0\left(\xi\right)&=&
v(\xi)
-\frac{i\omega(i\omega+\kappa_1)}{\kappa_1^2\left(\xi-\xi_1\right)^2}+\frac{i\omega(p'-2i\omega)}{\kappa_1p(\xi-\xi_1)}
\,.\label{s0}
\end{eqnarray}
Differentiating (\ref{ec.aim}) with respect to $\xi$ iteratively, for the $(n+2)$th derivatives, we have     
\begin{equation}\label{ec.aim2}
\frac{d^{n+2}\chi(\xi)}{d\xi^{n+2}}=\lambda_n(\xi)\frac{d\chi(\xi)}{d\xi}+s_n(\xi)\chi(\xi)\,,
\end{equation}
where the coefficients are
\begin{eqnarray}
	\lambda_n(\xi)&=&\frac{d\lambda_{n-1}(\xi)}{d\xi}+s_{n-1}(\xi)+\lambda_0(\xi)\lambda_{n-1}(\xi)\,,\label{ec.ln0}\\
	s_n(\xi)&=&\frac{ds_{n-1}(\xi)}{d\xi}+s_0(\xi)\lambda_{n-1}(\xi)\,.\label{ec.sn0}
\end{eqnarray}
For sufficiently large $n$, the
coefficients $\lambda_n(\xi)$ and $s_n(\xi)$ satisfy the quantization relation
\begin{equation}\label{ec.qz}
	s_n\lambda_{n-1}-s_{n-1}\lambda_n=0\,.	
\end{equation}
Now, expand the $\lambda_n(\xi)$ and $s_n(\xi)$ in a Taylor series around the point $\xi_0$, we have
\begin{eqnarray}
	\lambda_n(\xi)&=&\sum_{i=0}^{\infty}c_n^i\left(\xi-\xi_0\right)^i\,,\label{ec.ln}\\
	s_n(\xi)&=&\sum_{i=0}^{\infty}d_n^i\left(\xi-\xi_0\right)^i\,,\label{ec.sn}
\end{eqnarray} 
where $c_n^i$ and $d_n^i$ are the $i$th Taylor coefficients. Substituting (\ref{ec.ln}) and (\ref{ec.sn}) in (\ref{ec.ln0}) and (\ref{ec.sn0}), we have the following recursion relations for the coefficients
\begin{eqnarray}
	c_n^i&=&(i+1)c_{n-1}^{i+1}+d_{n-1}^i+\sum_{j=0}^ic_0^jc_{n-1}^{i-j}\,,\label{ec.ln1}\\
	d_n^i&=&(i+1)d_{n-1}^{i+1}+\sum_{j=0}^id_0^jc_{n-1}^{i-j}\,,\label{ec.sn1}
\end{eqnarray} 
then, the quantization condition (\ref{ec.qz}) can be rewritten as
\begin{equation}\label{ec.qz1}
	d_n^0c_{n-1}^0-d_{n-1}^0c_n^0=0\,.
\end{equation}
Using the quantization condition (\ref{ec.qz1}) we can obtain quasi-normal modes, for $n$ large enough.
\subsection{Results}

We have obtained numerically the quasi-normal modes for scalar, electromagnetic and gravitational perturbations by using the sixth-order WKB approach and AIM technique for Hayward, Bardeen and ABG BHs immersed in PFDM. The results are presented in Tables \ref{t1}-\ref{t2} for the scalar field, in Tables \ref{t3}-\ref{t4} for the electromagnetic and in Tables \ref{t5}-\ref{t6} for the gravitational perturbation.

We do not observe  noticeable differences in the computed quasi-normal mode frequencies between the WKB approximation and the AIM, except in the fundamental mode case where $n =l $ which presents differences in all cases (scalar, electromagnetic and gravitational perturbations). This discrepancy is expected, as it is well established that the WKB method becomes less accurate for low values of $l$, due to the limitations in its semi-classical expansion. However, as the angular momentum quantum number $l$ increases, the results obtained from both methods tend to converge, and the differences become significantly smaller.

\begin{table}[th]
	\begin{center}
	\resizebox{!}{.19\textwidth}{
		\begin{tabular}{c c c c c c c c c}
			\hline
			\hline
			\multicolumn{9}{c}{Scalar perturbations}\\ \hline
		&&&\multicolumn{2}{c}{Hayward}&\multicolumn{2}{c}{Bardeen}&\multicolumn{2}{c}{ABG}\\
		$n$&$l$\,&$\alpha$\,&WKB&AIM&WKB&AIM&WKB& AIM\\
			\multirow{4}{*}{0}&\multirow{2}{*}{1\,}&-0.1\,&
			0.534548-0.159678$i$
			\,	
			&
			0.534754-0.159095$i$
			\,
			&
			0.523641-0.151158$i$
			\,
			&
			0.523686-0.150930$i$
			\,
			&
			0.505599-0.139672$i$
			\,
			&
			0.505617-0.139515$i$
			\,
			\\
			&&-0.4\,&
			0.460480-0.115943$i$
			\,
			&
			0.460462-0.115885$i$
			\,
			&
			0.445778-0.107986$i$
			\,
			&
			0.445779-0.107929$i$
			\,
			&
			0.425937-0.098812$i$
			\,	
			&
			0.425927-0.098778$i$
			\,
			\\
			&\multirow{2}{*}{2\,}&-0.1\,&
		0.883211-0.158210$i$
		\,
		&
		0.883224-0.158165$i$
		\,	
		&
		0.864582-0.149975$i$
		\,	
		&
		0.864589-0.149951$i$
		\,
		&
		0.834947-0.138702$i$
		\,	
		&
		0.834952-0.138684$i$
		\,
		\\
			&&-0.4\,&
			0.763757-0.115586$i$
			\,
			&
			0.763757-0.115581$i$
			\,	
			&
			0.739371-0.107592$i$
			\,	
			&
			0.739373-0.107586$i$
			\,
			&
			0.706969-0.098460$i$
			\,	
			&
			0.706970-0.098455$i$
			\,
			\\
		\hline
			\multirow{6}{*}{1}&\multirow{2}{*}{1\,}&-0.1\,&
			0.493228-0.495366$i$
			\,
			&
			0.492926-0.492864$i$
			\,	
			&
			0.488685-0.467721$i$
			\,
			&
			0.488427-0.466973$i$
			\,
			&
			0.475128-0.430638$i$
			\,	
			&
			0.474917-0.430146$i$
			\,
			\\
			&&-0.4\,&
			0.434968-0.355180$i$
			\,
			&
			0.434834-0.354996$i$
			\,	
			&
			0.423197-0.330145$i$
			\,	
			&
			0.423143-0.329989$i$
			\,
			&
			0.404235-0.301475$i$
			\,	
			&
			0.404151-0.301406$i$
			\,
			\\
			&\multirow{2}{*}{2\,}&-0.1\,&
			0.854900-0.480999$i$
			\,
			&
			0.854906-0.480745$i$
			\,	
			&
			0.840647-0.455662$i$
			\,	
			&
			0.840647-0.455552$i$
			\,
			&
			0.814300-0.420790$i$
			\,	
			&
			0.814297-0.420709$i$
			\,
			\\
			&&-0.4\,&
			0.747104-0.349669$i$
			\,
			&
			0.747100-0.349639$i$
			\,	
			&
			0.724818-0.325224$i$
			\,	
			&
			0.724821-0.325193$i$
			\,
			&
			0.693314-0.297337$i$
			\,	
			&
			0.693309-0.297316$i$
			\,
			\\
			&\multirow{2}{*}{3\,}&-0.1\,&
		1.212610-0.477094$i$
		\,
		&
		1.212610-0.477053$i$
		\,
		&
		1.189610-0.452043$i$
		\,	
		&
		1.189610-0.452023$i$
		\,
		&
		1.150750-0.417809$i$
		\,	
		&
		1.150750-0.417794$i$
		\,
		\\
			&&-0.4\,&
		1.055730-0.348037$i$
		\,
		&
		1.055730-0.348033$i$
		\,	
		&
		1.023160-0.323756$i$
		\,
		&
		1.023160-0.323751$i$
		\,
		&
		0.978761-0.296116$i$
		\,	
		&
		0.978761-0.296112$i$
		\,
		\\
		\hline
			\multirow{6}{*}{2}&\multirow{2}{*}{2\,}&-0.1\,&
			0.805244-0.820682$i$
			\,	
			&
			0.805116-0.819916$i$
			\,	
			&
			0.799224-0.776556$i$
			\,	
			&
			0.799120-0.776206$i$
			\,
			&
			0.778204-0.715287$i$
			\,	
			&
			0.778095-0.715026$i$
			\,
			\\
			&&-0.4\,&
			0.716529-0.591648$i$
			\,
			&
			0.716493-0.591526$i$
			\,	
			&
			0.697942-0.549516$i$
			\,
			&
			0.697922-0.549402$i$
			\,
			&
			0.667455-0.501674$i$
			\,	
			&
			0.667406-0.501609$i$
			\,
			\\
			&\multirow{2}{*}{3\,}&-0.1\,&
			1.173380-0.805597$i$
			\,
			&
			1.173370-0.805470$i$
			\,
			&
			1.156670-0.762884$i$
			\,	
			&
			1.156660-0.762824$i$
			\,
			&
			1.122270-0.704088$i$
			\,	
			&
			1.122260-0.704039$i$
			\,
			\\
			&&-0.4\,&
			1.032430-0.584896$i$
			\,	
			&
			1.032430-0.584877$i$
			\,	
			&
			1.002820-0.543658$i$
			\,	
			&
			1.002810-0.543640$i$
			\,
			&
			0.959546-0.496798$i$
			\,	
			&
			0.959540-0.496785$i$
			\,
			\\
			&\multirow{2}{*}{4\,}&-0.1\,&
			1.536160-0.799012$i$
			\,
			&
			1.536160-0.798981$i$
			\,
			&
			1.509910-0.756804$i$
			\,
			&
			1.509910-0.756789$i$
			\,
			&
			1.462500-0.699100$i$
			\,	
			&
			1.462500-0.699088$i$
			\,
			\\
			&&-0.4\,&
		1.344140-0.581933$i$
		\,
		&
		1.344140-0.581929$i$
		\,	
		&
		1.303880-0.541098$i$
		\,	
		&
		1.303880-0.541093$i$
		\,
		&
		1.247600-0.494700$i$
		\,
		&
		1.247600-0.494695$i$
		\,
		\\
			\hline
			\hline		\end{tabular}
		}
		\caption{Quasi-normal frequencies for the scalar perturbations for several values of the parameter $\alpha$, with $q_1^2=q_2^2=q_3^2=0.1$.}
		\label{t1}
	\end{center}
\end{table}

\begin{table}[th]
	\begin{center}
		\resizebox{!}{.19\textwidth}{
		\begin{tabular}{c c c c c c c c c}
			\hline
			\hline
			\multicolumn{9}{c}{Scalar perturbations}\\ \hline
			&&&\multicolumn{2}{c}{Hayward}&\multicolumn{2}{c}{Bardeen}&\multicolumn{2}{c}{ABG}\\
			$n$&$l$\,\,&$q^2_i$\,\,&WKB&AIM&WKB&AIM&WKB& AIM\\
			\multirow{4}{*}{0}&\multirow{2}{*}{1\,\,}&0.1\,\,&
			0.534548-0.159678$i$
			\,\,	
			&
			0.534754-0.159095$i$
			\,\,	
			&
			0.523641-0.151158$i$
			\,\,	
			&
			0.523686-0.150930$i$
			\,\,
			&
			0.505599-0.139672$i$
			\,\,	
			&
			0.505617-0.139515$i$
			\,\,
			\\
			&&0.3\,\,&
			0.469992-0.126522$i$
			\,\,	
			&
			0.470500-0.125543$i$
			\,\,	
			&
			0.450732-0.111660$i$
			\,\,	
			&
			0.450756-0.111483$i$
			\,\,
			&
			0.406760-0.093135$i$
			\,\,	
			&
			0.406762-0.093032$i$
			\,\,
			\\
			&\multirow{2}{*}{2\,\,}&0.1\,\,&
			0.883211-0.158210$i$
			\,\,	
			&
			0.883224-0.158165$i$
			\,\,	
			&
			0.864582-0.149975$i$
			\,\,	
			&
			0.864589-0.149951$i$
			\,\,
			&
			0.834947-0.138702$i$
			\,\,	
			&
			0.834952-0.138684$i$
			\,\,
			\\
			&&0.3\,\,&
			0.778835-0.125238$i$
			\,\,	
			&
			0.778891-0.125129$i$
			\,\,	
			&
			0.745463-0.110877$i$
			\,\,	
			&
			0.745471-0.110854$i$
			\,\,
			&
			0.674227-0.092397$i$
			\,\,	
			&
			0.674228-0.092385$i$
			\,\,
			\\
			\hline
			\multirow{6}{*}{1}&\multirow{2}{*}{1\,\,}&0.1\,\,&
			0.493228-0.495366$i$
			\,\,	
			&
	        0.492926-0.492864$i$
			\,\,	
			&
			0.488685-0.467721$i$
			\,\,	
			&
			0.488427-0.466973$i$
			\,\,
			&
			0.475128-0.430638$i$
			\,\,	
			&
			0.474917-0.430146$i$
			\,\,
			\\
			&&0.3\,\,&
		0.426200-0.390212$i$
		\,\,	
		&
		0.426359-0.385741$i$
		\,\,	
		&
		0.421544-0.341285$i$
		\,\,	
		&
		0.421361-0.340705$i$
		\,\,
		&
		0.379300-0.284582$i$
		\,\,	
		&
		0.379126-0.284278$i$
		\,\,
		\\
			&\multirow{2}{*}{2\,\,}&0.1\,\,&
		0.854900-0.480999$i$
		\,\,	
		&
		0.854906-0.480745$i$
		\,\,	
		&
		0.840647-0.455662$i$
		\,\,	
		&
		0.840647-0.455552$i$
		\,\,
		&
		0.814300-0.420790$i$
		\,\,	
		&
		0.814297-0.420709$i$
		\,\,
		\\
			&&0.3\,\,&
			0.742573-0.433324$i$
			\,\,	
			&
			0.751518-0.378696$i$
			\,\,	
			&
			0.727113-0.335060$i$
			\,\,	
			&
			0.727121-0.334947$i$
			\,\,
			&
			0.657446-0.278999$i$
			\,\,	
			&
			0.657422-0.278945$i$
			\,\,
			\\
			&\multirow{2}{*}{3\,\,}&0.1\,\,&
		1.212610-0.477094$i$
		\,\,	
		&
		1.212610-0.477053$i$
		\,\,	
		&
		1.189610-0.452043$i$
		\,\,	
		&
		1.189610-0.452023$i$
		\,\,
		&
		1.150750-0.417809$i$
		\,\,	
		&
		1.150750-0.417794$i$
		\,\,
		\\
			&&0.3\,\,&
		1.068710-0.376943$i$
		\,\,	
		&
		1.068740-0.376833$i$
		\,\,	
		&
		1.028140-0.333275$i$
		\,\,	
		&
		1.028150-0.333255$i$
		\,\,
		&
		0.930408-0.277513$i$
		\,\,	
		&
		0.930409-0.277501$i$
		\,\,
		\\
			\hline
			\multirow{6}{*}{2}&\multirow{2}{*}{2\,\,}&0.1\,\,&
		0.805244-0.820682$i$
		\,\,	
		&
		0.805116-0.819916$i$
		\,\,	
		&
		0.799224-0.776556$i$
		\,\,	
		&
		0.799120-0.776206$i$
		\,\,
		&
		0.778204-0.715287$i$
		\,\,	
		&
		0.778095-0.715026$i$
		\,\,
		\\
			&&0.3\,\,&
			0.697747-0.644039$i$
			\,\,	
			&
			0.697668-0.642337$i$
			\,\,	
			&
			0.692032-0.566023$i$
			\,\,	
			&
			0.691959-0.565690$i$
			\,\,
			&
			0.624343-0.471061$i$
			\,\,	
			&
			0.624222-0.471018$i$
			\,\,
			\\
			&\multirow{2}{*}{3\,\,}&0.1\,\,&
		1.173380-0.805597$i$
		\,\,	
		&
		1.173370-0.805470$i$
		\,\,	
		&
		1.156670-0.762884$i$
		\,\,	
		&
		1.156660-0.762824$i$
		\,\,
		&
		1.122270-0.704088$i$
		\,\,	
		&
		1.122260-0.704039$i$
		\,\,
		\\
			&&0.3\,\,&
			1.029510-0.634045$i$
			\,\,	
			&
			1.029530-0.633708$i$
			\,\,	
			&
			1.002240-0.559472$i$
			\,\,	
			&
			1.002230-0.559410$i$
			\,\,
			&
			0.906462-0.465573$i$
			\,\,	
			&
			0.906461-0.465541$i$
			\,\,
			\\
			&\multirow{2}{*}{4\,\,}&0.1\,\,&
		1.536160-0.799012$i$
		\,\,	
		&
		1.536160-0.798981$i$
		\,\,	
		&
		1.509910-0.756804$i$
		\,\,	
		&
		1.509910-0.756789$i$
		\,\,
		&
		1.462500-0.699100$i$
		\,\,	
		&
		1.462500-0.699088$i$
		\,\,
		\\
			&&0.3\,\,&
		1.352290-0.630314$i$
		\,\,	
		&
		1.352300-0.630225$i$
		\,\,	
		&
		1.306960-0.556717$i$
		\,\,	
		&
		1.306960-0.556701$i$
		\,\,
		&
		1.182870-0.463347$i$
		\,\,	
		&
		1.182870-0.463338$i$
		\,\,
		\\
			\hline
			\hline		\end{tabular}
		}
		\caption{Quasi-normal frequencies for the scalar perturbations for several values of the parameter $q^2_i$, with $\alpha=-0.1$.}
		\label{t2}
	\end{center}
\end{table}

\begin{table}[th]
	\begin{center}
		\resizebox{!}{.19\textwidth}{
			\begin{tabular}{c c c c c c c c c}
				\hline
				\hline
				\multicolumn{9}{c}{Electromagnetic perturbations}\\ \hline
				&&&\multicolumn{2}{c}{Hayward}&\multicolumn{2}{c}{Bardeen}&\multicolumn{2}{c}{ABG}\\
				$n$&$l$\,&$\alpha$\,&WKB&AIM&WKB&AIM&WKB& AIM\\
				\multirow{4}{*}{0}&\multirow{2}{*}{1\,}&-0.1\,&
				0.461792-0.152884$i$
				\,	
				&
				0.462280-0.151658$i$
				\,
				&
				0.454534-0.145260$i$
				\,
				&
				0.454850-0.144544$i$
				\,
				&
				0.441624-0.134512$i$
				\,
				&
				0.441881-0.133937$i$
				\,
				\\
				&&-0.4\,&
				0.409281-0.112209$i$
				\,	
				&
				0.409324-0.111979$i$
				\,
				&
				0.397697-0.104570$i$
				\,
				&
				0.397782-0.104343$i$
				\,
				&
				0.381588-0.095501$i$
				\,
				&
				0.381651-0.095319$i$
				\,
				\\
				&\multirow{2}{*}{2\,}&-0.1\,&
				0.840541-0.155561$i$
				\,	
				&
				0.840557-0.155506$i$
				\,
				&
				0.824108-0.147709$i$
				\,
				&
				0.824119-0.147676$i$
				\,
				&
				0.797406-0.136708$i$
				\,
				&
				0.797414-0.136682$i$
				\,
				\\
				&&-0.4\,&
				0.733630-0.114205$i$
				\,	
				&
				0.733630-0.114196$i$
				\,
				&
				0.711056-0.106315$i$
				\,
				&
				0.711059-0.106306$i$
				\,
				&
				0.680796-0.097229$i$
				\,
				&
				0.680798-0.0972210$i$
				\,
				\\
				\hline
				\multirow{6}{*}{1}&\multirow{2}{*}{1\,}&-0.1\,&
				0.415747-0.479891$i$
				\,	
				&
				0.415319-0.473850$i$
				\,
				&
				0.414644-0.453896$i$
				\,
				&
				0.414492-0.450846$i$
				\,
				&
				0.407197-0.418337$i$
				\,
				&
				0.407135-0.415887$i$
				\,
				\\
				&&-0.4\,&
				0.381582-0.345622$i$
				\,	
				&
				0.381413-0.344639$i$
				\,
				&
				0.373354-0.321240$i$
				\,
				&
				0.373331-0.320306$i$
				\,
				&
				0.358588-0.292356$i$
				\,
				&
				0.358529-0.291638$i$
				\,
				\\
				&\multirow{2}{*}{2\,}&-0.1\,&
				0.811069-0.473524$i$
				\,	
				&
				0.811074-0.473204$i$
				\,
				&
				0.799009-0.449284$i$
				\,
				&
				0.799015-0.449124$i$
				\,
				&
				0.775836-0.415151$i$
				\,
				&
				0.775842-0.415026$i$
				\,
				\\
				&&-0.4\,&
				0.716445-0.345705$i$
				\,	
				&
				0.716437-0.345659$i$
				\,
				&
				0.696086-0.321540$i$
				\,
				&
				0.696089-0.321492$i$
				\,
				&
				0.666827-0.293736$i$
				\,
				&
				0.666833-0.293696$i$
				\,
				\\
				&\multirow{2}{*}{3\,}&-0.1\,&
				1.181850-0.473139$i$
				\,	
				&
				1.181850-0.473093$i$
				\,
				&
				1.160420-0.448665$i$
				\,
				&
				1.160420-0.448643$i$
				\,
				&
				1.123720-0.414825$i$
				\,
				&
				1.123720-0.414807$i$
				\,
				\\
				&&-0.4\,&
				1.034110-0.345972$i$
				\,	
				&
				1.034110-0.345966$i$
				\,
				&
				1.002870-0.321837$i$
				\,
				&
				1.002870-0.321831$i$
				\,
				&
				0.960024-0.294253$i$
				\,
				&
				0.960025-0.294248$i$
				\,
				\\
				\hline
				\multirow{6}{*}{2}&\multirow{2}{*}{2\,}&-0.1\,&
				0.759660-0.809638$i$
				\,	
				&
				0.759527-0.808670$i$
				\,
				&
				0.755771-0.767180$i$
				\,
				&
				0.755682-0.766682$i$
				\,
				&
				0.738323-0.706908$i$
				\,
				&
				0.738229-0.706510$i$
				\,
				\\
				&&-0.4\,&
				0.685061-0.585557$i$
				\,	
				&
				0.684996-0.585403$i$
				\,
				&
				0.668580-0.543791$i$
				\,
				&
				0.668548-0.543637$i$
				\,
				&
				0.640451-0.495940$i$
				\,
				&
				0.640441-0.495802$i$
				\,
				\\
				&\multirow{2}{*}{3\,}&-0.1\,&
				1.141860-0.799380$i$
				\,	
				&
				1.141850-0.799240$i$
				\,
				&
				1.126720-0.757593$i$
				\,
				&
				1.126710-0.757523$i$
				\,
				&
				1.094640-0.699386$i$
				\,
				&
				1.094640-0.699329$i$
				\,
				\\
				&&-0.4\,&
				1.010460-0.581598$i$
				\,	
				&
				1.010450-0.581577$i$
				\,
				&
				0.982249-0.540576$i$
				\,
				&
				0.982246-0.540554$i$
				\,
				&
				0.940602-0.493765$i$
				\,
				&
				0.940603-0.493745$i$
				\,
				\\
				&\multirow{2}{*}{4\,}&-0.1\,&
				1.512020-0.795111$i$
				\,	
				&
				1.512020-0.795078$i$
				\,
				&
				1.487000-0.753477$i$
				\,
				&
				1.487000-0.753462$i$
				\,
				&
				1.441320-0.696153$i$
				\,
				&
				1.441320-0.696140$i$
				\,
				\\
				&&-0.4\,&
				1.327240-0.579890$i$
				\,	
				&
				1.327240-0.579884$i$
				\,
				&
				1.288040-0.539192$i$
				\,
				&
				1.288040-0.539186$i$
				\,
				&
				1.232990-0.492837$i$
				\,
				&
				1.232990-0.492834$i$
				\,
				\\
				\hline
				\hline		
			\end{tabular}}
		\caption{Quasi-normal frequencies for the electromagnetic perturbations for several values of the parameter $\alpha$, with $q_1^2=q_2^2=q_3^2=0.1$.}
		\label{t3}
	\end{center}
\end{table}

\begin{table}[th]
	\begin{center}
			\resizebox{!}{.19\textwidth}{
		\begin{tabular}{c c c c c c c c c}
			\hline
			\hline
			\multicolumn{9}{c}{Electromagnetic perturbations}\\ \hline
			&&&\multicolumn{2}{c}{Hayward}&\multicolumn{2}{c}{Bardeen}&\multicolumn{2}{c}{ABG}\\
			$n$&$l$\,\,&$q_i^2$\,\,&WKB&AIM&WKB&AIM&WKB& AIM\\
		\multirow{4}{*}{0}&\multirow{2}{*}{1\,\,}&0.1\,\,&
		0.461792-0.152884$i$
		\,\,	
		&
		0.462280-0.151658$i$
		\,\,	
		&
		0.454534-0.145260$i$
		\,\,	
		&
		0.454850-0.144544$i$
		\,\,
		&
		0.441624-0.134512$i$
		\,\,	
		&
		0.441881-0.133937$i$
		\,\,
		\\
		&&0.3\,\,&
		0.409793-0.120846$i$
		\,\,	
		&
		0.411812-0.118010$i$
		\,\,	
		&
		0.398150-0.106830$i$
		\,\,	
		&
		0.398520-0.106121$i$
		\,\,
		&
		0.362592-0.088619$i$
		\,\,	
		&
		0.362767-0.088226$i$
		\,\,
		\\
		&\multirow{2}{*}{2\,\,}&0.1\,\,&
		0.840541-0.155561$i$
		\,\,	
		&
		0.840557-0.155506$i$
		\,\,	
		&
		0.824108-0.147709$i$
		\,\,	
		&
		0.824119-0.147676$i$
		\,\,
		&
		0.797406-0.136708$i$
		\,\,	
		&
		0.797414-0.136682$i$
		\,\,
		\\
		&&0.3\,\,&
		0.743698-0.122510$i$
		\,\,	
		&
		0.743780-0.122364$i$
		\,\,	
		&
		0.714356-0.108916$i$
		\,\,	
		&
		0.714369-0.108883$i$
		\,\,
		&
		0.648009-0.090655$i$
		\,\,	
		&
		0.648015-0.090637$i$
		\,\,
		\\
		\hline
		\multirow{6}{*}{1}&\multirow{2}{*}{1\,\,}&0.1\,\,&
		0.415747-0.479891$i$
		\,\,	
		&
		0.415319-0.473850$i$
		\,\,	
		&
		0.414645-0.453895$i$
		\,\,	
		&
		0.414492-0.450846$i$
		\,\,
		&
		0.407197-0.418337$i$
		\,\,	
		&
		0.407135-0.415887$i$
		\,\,
		\\
		&&0.3\,\,&
		0.363736-0.378711$i$
		\,\,	
		&
		0.366278-0.363027$i$
		\,\,	
		&
		0.367871-0.328506$i$
		\,\,	
		&
		0.368104-0.325165$i$
		\,\,
		&
		0.334045-0.271718$i$
		\,\,	
		&
		0.334103-0.269785$i$
		\,\,
		\\
		&\multirow{2}{*}{2\,\,}&0.1\,\,&
		0.811069-0.473524$i$
		\,\,	
		&
		0.811074-0.473204$i$
		\,\,	
		&
		0.799009-0.449284$i$
		\,\,	
		&
		0.799015-0.449124$i$
		\,\,
		&
		0.775836-0.415151$i$
		\,\,	
		&
		0.775842-0.415026$i$
		\,\,
		\\
		&&0.3\,\,&
		0.715968-0.371277$i$
		\,\,	
		&
		0.716190-0.370411$i$
		\,\,	
		&
		0.695803-0.329286$i$
		\,\,	
		&
		0.695824-0.329114$i$
		\,\,
		&
		0.631052-0.273797$i$
		\,\,	
		&
		0.631073-0.273695$i$
		\,\,
		\\
		&\multirow{2}{*}{3\,\,}&0.1\,\,&
		1.181850-0.473139$i$
		\,\,	
		&
		1.181850-0.473093$i$
		\,\,	
		&
		1.160420-0.448665$i$
		\,\,	
		&
		1.160420-0.448643$i$
		\,\,
		&
		1.123720-0.414825$i$
		\,\,	
		&
		1.123720-0.414807$i$
		\,\,
		\\
		&&0.3\,\,&
		1.043560-0.372698$i$
		\,\,	
		&
		1.043590-0.372570$i$
		\,\,	
		&
		1.005890-0.330257$i$
		\,\,	
		&
		1.005900-0.330232$i$
		\,\,
		&
		0.911666-0.274824$i$
		\,\,	
		&
		0.911669-0.274810$i$
		\,\,
		\\
		\hline
		\multirow{6}{*}{2}&\multirow{2}{*}{2\,\,}&0.1\,\,&
		0.759660-0.809638$i$
		\,\,	
		&
		0.759527-0.808670$i$
		\,\,	
		&
		0.755771-0.767180$i$
		\,\,	
		&
		0.755682-0.766682$i$
		\,\,
		&
		0.738323-0.706908$i$
		\,\,	
		&
		0.738229-0.706510$i$
		\,\,
		\\
		&&0.3\,\,&
		0.661940-0.631013$i$
		\,\,	
		&
		0.661932-0.628489$i$
		\,\,	
		&
		0.660514-0.556669$i$
		\,\,	
		&
		0.660437-0.556167$i$
		\,\,
		&
		0.597510-0.462521$i$
		\,\,	
		&
		0.597575-0.462259$i$
		\,\,
		\\
		&\multirow{2}{*}{3\,\,}&0.1\,\,&
		1.141860-0.799380$i$
		\,\,	
		&
		1.141850-0.799240$i$
		\,\,	
		&
		1.126720-0.757593$i$
		\,\,	
		&
		1.126710-0.757523$i$
		\,\,
		&
		1.094640-0.699386$i$
		\,\,	
		&
		1.094640-0.699329$i$
		\,\,
		\\
		&&0.3\,\,&
		1.004240-0.627012$i$
		\,\,	
		&
		1.004260-0.626613$i$
		\,\,	
		&
		0.979891-0.554513$i$
		\,\,	
		&
		0.979884-0.554438$i$
		\,\,
		&
		0.887627-0.461095$i$
		\,\,	
		&
		0.887640-0.461050$i$
		\,\,
		\\
		&\multirow{2}{*}{4\,\,}&0.1\,\,&
		1.512020-0.795111$i$
		\,\,	
		&
		1.512020-0.795078$i$
		\,\,	
		&
		1.487000-0.753477$i$
		\,\,	
		&
		1.487000-0.753462$i$
		\,\,
		&
		1.441320-0.696153$i$
		\,\,	
		&
		1.441320-0.696140$i$
		\,\,
		\\
		&&0.3\,\,&
		1.332700-0.626013$i$
		\,\,	
		&
		1.332710-0.625916$i$
		\,\,	
		&
		1.289630-0.553676$i$
		\,\,	
		&
		1.289630-0.553658$i$
		\,\,
		&
		1.168280-0.460625$i$
		\,\,	
		&
		1.168280-0.460616$i$
		\,\,
		\\
			\hline
			\hline		\end{tabular}}
		\caption{Quasi-normal frequencies for the scalar perturbations for several values of the parameter $q_i^2$, with $\alpha=-0.1$.}
		\label{t4}
	\end{center}
\end{table}

\begin{table}[th]
	\begin{center}
		\resizebox{!}{.19\textwidth}{
			\begin{tabular}{c c c c c c c c c}
				\hline
				\hline
				\multicolumn{9}{c}{Gravitational perturbations}\\ \hline
				&&&\multicolumn{2}{c}{Hayward}&\multicolumn{2}{c}{Bardeen}&\multicolumn{2}{c}{ABG}\\
				$n$&$l$\,&$\alpha$\,&WKB&AIM&WKB&AIM&WKB& AIM\\
				\multirow{4}{*}{0}&\multirow{2}{*}{2\,}&-0.1\,&
0.686178-0.145713$i$
				\,	
				&
0.685971-0.145703$i$
				\,				
				&
0.672485-0.138624$i$				
				\,
				&
0.672528-0.138486$i$				
				\,
	     		&
0.650533-0.128428$i$		
		        \,
		        &
0.650631-0.128260$i$
				\,
				\\
				&&-0.4\,&
0.598034-0.107722$i$
				\,	
				&
0.598043-0.107672$i$
				\,
				&
0.579616-0.100354$i$
				\,
				&
0.579662-0.100285$i$					
				\,
				&
0.554891-0.091818$i$				
				\,
				&
0.554930-0.091759$i$				
				\,
				\\
				&\multirow{2}{*}{3\,}&-0.1\,&
1.09781-0.152084$i$	
				\,	
				&
1.097810-0.152073$i$
				\,
				&
1.075320-0.144363$i$				
				\,
				&
1.075320-0.144355$i$				
				\,
				&
1.039490-0.133683$i$		
				\,
				&
1.039490-0.133677$i$				
				\,
				\\
				&&-0.4\,&
0.954711-0.112006$i$				
				\,	
				&
0.954711-0.112004$i$
				\,
				&
0.9247510-0.10429$i$				
				\,
				&
0.924751-0.104287$i$				
				\,
				&
0.884962-0.095453$i$
				\,
				&
0.884962-0.095451$i$			
				\,
				\\
				\hline
				\multirow{6}{*}{1}&\multirow{2}{*}{2\,}&-0.1\,&
0.650442-0.445727$i$
				\,	
				&
0.649380-0.445628$i$
				\,
				&
0.641119-0.424410$i$
				\,
				&
0.641183-0.423401$i$
				\,
				&
0.623561-0.392587$i$
				\,
				&
0.623875-0.391361$i$
				\,
				\\
				&&-0.4\,&
0.576838-0.327510$i$
				\,	
				&
0.576855-0.327101$i$
				\,
				&
0.561166-0.304836$i$
				\,
				&
0.561344-0.304339$i$
				\,
				&
0.537979-0.278416$i$
				\,
				&
0.538117-0.278023$i$
				\,
				\\
				&\multirow{2}{*}{3\,}&-0.1\,&
1.074310-0.460118$i$
				\,	
				&
1.074320-0.460037$i$
				\,
				&
1.055270-0.436620$i$
				\,
				&
1.055270-0.436577$i$
				\,
				&
1.022260-0.403952$i$
				\,
				&
1.022270-0.403919$i$
				\,
				\\
				&&-0.4\,&
0.941033-0.337818$i$
				\,	
				&
0.941033-0.337807$i$
				\,
				&
0.912874-0.314388$i$
				\,
				&
0.912874-0.314375$i$
				\,
				&
0.873997-0.287560$i$
				\,
				&
0.873993-0.287552$i$
				\,
				\\
				&\multirow{2}{*}{4\,}&-0.1\,&
1.462780-0.465524$i$
				\,	
				&
1.462780-0.465511$i$
				\,
				&
1.434510-0.441494$i$
				\,
				&
1.434510-0.441487$i$
				\,
				&
1.387830-0.408484$i$
				\,
				&
1.387830-0.408478$i$
				\,
				\\
				&&-0.4\,&
1.275560-0.341474$i$
				\,	
				&
1.275560-0.341472$i$
				\,
				&
1.236220-0.317767$i$
				\,
				&
1.236220-0.317765$i$
				\,
				&
1.183100-0.290720$i$
				\,
				&
1.183100-0.290718$i$
				\,
				\\
				\hline
				\multirow{6}{*}{2}&\multirow{2}{*}{2\,}&-0.1\,&
0.588875-0.768549$i$
				\,	
				&
0.586111-0.768542$i$
				\,
				&
0.587955-0.733069$i$
				\,
				&
0.587761-0.729564$i$
				\,
				&
0.577717-0.676303$i$
				\,
				&
0.577974-0.671969$i$
				\,
				\\
				&&-0.4\,&
0.538828-0.559088$i$
				\,	
				&
0.538668-0.557535$i$
				\,
				&
0.527899-0.519598$i$
				\,
				&
0.528135-0.517794$i$
				\,
				&
0.506585-0.473231$i$
				\,
				&
0.506647-0.471926$i$
				\,
				\\
				&\multirow{2}{*}{3\,}&-0.1\,&
1.030380-0.778906$i$
				\,	
				&
1.030380-0.778906$i$
				\,
				&
1.018040-0.738907$i$
				\,
				&
1.018050-0.738781$i$
				\,
				&
0.990178-0.682441$i$
				\,
				&
0.990176-0.682344$i$
				\,
				\\
				&&-0.4\,&
0.914949-0.568768$i$
				\,	
				&
0.914950-0.568723$i$
				\,
				&
0.890176-0.528817$i$
				\,
				&
0.890168-0.528780$i$
				\,
				&
0.852764-0.483118$i$
				\,
				&
0.852732-0.483101$i$
				\,
				\\
				&\multirow{2}{*}{4\,}&-0.1\,&
1.428780-0.783008$i$
				\,	
				&
1.428780-0.782968$i$
				\,
				&
1.405720-0.742345$i$
				\,
				&
1.405720-0.742325$i$
				\,
				&
1.363030-0.686150$i$
				\,
				&
1.363040-0.686131$i$
				\,
				\\
				&&-0.4\,&
1.255620-0.572445$i$
				\,	
				&
1.255620-0.572439$i$
				\,
				&
1.218880-0.532410$i$
				\,
				&
1.218880-0.532403$i$
				\,
				&
1.166930-0.486761$i$
				\,
				&
1.166930-0.486754$i$
				\,
				\\
				\hline
				\hline		
		\end{tabular}}
		\caption{Quasinormal frequencies for the gravitational perturbations for several values of the parameter $\alpha$, with $q_1^2=q_2^2=q_3^2=0.1$.}
		\label{t5}
	\end{center}
\end{table}

\begin{table}[th]
	\begin{center}
		\resizebox{!}{.19\textwidth}{
			\begin{tabular}{c c c c c c c c c}
				\hline
				\hline
				\multicolumn{9}{c}{Gravitational perturbations}\\ \hline
				&&&\multicolumn{2}{c}{Hayward}&\multicolumn{2}{c}{Bardeen}&\multicolumn{2}{c}{ABG}\\
				$n$&$l$\,\,&$q_i^2$\,\,&WKB&AIM&WKB&AIM&WKB& AIM\\
				\multirow{4}{*}{0}&\multirow{2}{*}{2\,\,}&0.1\,\,&
0.686178-0.145713$i$
				\,\,	
				&
0.685971-0.145703$i$
				\,\,	
				&
0.672485-0.138624$i$
				\,\,	
				&
0.672528-0.138486$i$
				\,\,
				&
0.650533-0.128428$i$
				\,\,	
				&
0.650631-0.128260$i$
				\,\,
				\\
				&&0.3\,\,&
0.605705-0.115537$i$
				\,\,	
				&
0.606559-0.114448$i$
				\,\,	
				&
0.582337-0.102441$i$
				\,\,	
				&
0.582640-0.102133$i$
				\,\,
				&
0.528182-0.085273$i$
				\,\,	
				&
0.528277-0.085167$i$
				\,\,
				\\
				&\multirow{2}{*}{3\,\,}&0.1\,\,&
1.097810-0.152084$i$
				\,\,	
				&
1.097810-0.152073$i$
				\,\,	
				&
1.075320-0.144363$i$
				\,\,	
				&
1.075320-0.144355$i$
				\,\,
				&
1.039490-0.133683$i$
				\,\,	
				&
1.039490-0.133677$i$
				\,\,
				\\
				&&0.3\,\,&
0.970126-0.120169$i$
				\,\,	
				&
0.970147-0.120131$i$
				\,\,	
				&
0.929876-0.106766$i$
				\,\,	
				&
0.929878-0.106758$i$
				\,\,
				&
0.842728-0.089014$i$
				\,\,	
				&
0.842729-0.089009$i$
				\,\,
				\\
				\hline
				\multirow{6}{*}{1}&\multirow{2}{*}{2\,\,}&0.1\,\,&
0.650442-0.445727$i$
				\,\,	
				&
0.649380-0.445628$i$
				\,\,	
				&
0.641119-0.424410$i$
				\,\,	
				&
0.641183-0.423401$i$
				\,\,
				&
0.623561-0.392587$i$
				\,\,	
				&
0.623875-0.391361$i$
				\,\,
				\\
				&&0.3\,\,&
0.571373-0.356296$i$
				\,\,	
				&
0.574539-0.347285$i$
				\,\,	
				&
0.559551-0.312246$i$
				\,\,	
				&
0.561019-0.309678$i$
				\,\,
				&
0.508433-0.258711$i$
				\,\,	
				&
0.508956-0.257878$i$
				\,\,
				\\
				&\multirow{2}{*}{3\,\,}&0.1\,\,&
1.074310-0.460118$i$
				\,\,	
				&
1.074320-0.460037$i$
				\,\,	
				&
1.055270-0.436620$i$
				\,\,	
				&
1.055270-0.436577$i$
				\,\,
				&
1.022260-0.403952$i$
				\,\,	
				&
1.022270-0.403919$i$
				\,\,
				\\
				&&0.3\,\,&
0.948829-0.362534$i$
				\,\,	
				&
0.948916-0.362283$i$
				\,\,	
				&
0.915486-0.321761$i$
				\,\,	
				&
0.915489-0.321722$i$
				\,\,
				&
0.829800-0.268080$i$
				\,\,	
				&
0.829804-0.268058$i$
				\,\,
				\\
				&\multirow{2}{*}{4\,\,}&0.1\,\,&
1.462780-0.465524$i$
				\,\,	
				&
1.462780-0.465511$i$
				\,\,	
				&
1.434510-0.441494$i$
				\,\,	
				&
1.434510-0.441487$i$
				\,\,
				&
1.387830-0.408484$i$
				\,\,	
				&
1.387830-0.408478$i$
				\,\,
				\\
				&&0.3\,\,&
1.291750-0.367707$i$
				\,\,	
				&
1.291760-0.367667$i$
				\,\,	
				&
1.241550-0.325917$i$
				\,\,	
				&
1.241550-0.325908$i$
				\,\,
				&
1.124930-0.271513$i$
				\,\,	
				&
1.124930-0.271509$i$
				\,\,
				\\
				\hline
				\multirow{6}{*}{2}&\multirow{2}{*}{2\,\,}&0.1\,\,&
0.588875-0.768549$i$
				\,\,	
				&
0.586111-0.768542$i$
				\,\,	
				&
0.587955-0.733069$i$
				\,\,	
				&
0.587761-0.729564$i$
				\,\,
				&
0.577717-0.676303$i$
				\,\,	
				&
0.577974-0.671969$i$
				\,\,
				\\
				&&0.3\,\,&
0.510825-0.624988$i$
				\,\,	
				&
0.513012-0.591845$i$
				\,\,	
				&
0.517058-0.537039$i$
				\,\,	
				&
0.520455-0.526720$i$
				\,\,
				&
0.469238-0.441818$i$
				\,\,	
				&
0.467887-0.435667$i$
				\,\,
				\\
				&\multirow{2}{*}{3\,\,}&0.1\,\,&
1.030380-0.778906$i$
				\,\,	
				&
1.030380-0.778906$i$
				\,\,	
				&
1.018040-0.738907$i$
				\,\,	
				&
1.018050-0.738781$i$
				\,\,
				&
0.990178-0.682441$i$
				\,\,	
				&
0.990176-0.682344$i$
				\,\,
				\\
				&&0.3\,\,&
0.906684-0.611014$i$
				\,\,	
				&
0.906825-0.610143$i$
				\,\,	
				&
0.887476-0.540967$i$
				\,\,	
				&
0.887456-0.540867$i$
				\,\,
				&
0.804086-0.450288$i$
				\,\,	
				&
0.804097-0.450239$i$
				\,\,
				\\
				&\multirow{2}{*}{4\,\,}&0.1\,\,&
1.428780-0.783008$i$
				\,\,	
				&
1.428780-0.782968$i$
				\,\,	
				&
1.405720-0.742345$i$
				\,\,	
				&
1.405720-0.742325$i$
				\,\,
				&
1.363030-0.686150$i$
				\,\,	
				&
1.363040-0.686131$i$
				\,\,
				\\
				&&0.3\,\,&
1.259650-0.616560$i$
				\,\,	
				&
1.259660-0.616431$i$
				\,\,	
				&
1.220090-0.545903$i$
				\,\,	
				&
1.220090-0.545877$i$
				\,\,
				&
1.105410-0.454494$i$
				\,\,	
				&
1.105400-0.454481$i$
				\,\,
				\\
				\hline
				\hline		\end{tabular}}
		\caption{Quasinormal frequencies for the gravitational perturbations for several values of the parameter $q_i^2$, with $\alpha=-0.1$.}
		\label{t6}
	\end{center}
\end{table}

Tables \ref{t1} and \ref{t3} show the quasi-normal mode frequencies for scalar and electromagnetic perturbations, respectively. It is observed that the real part of the frequency increases with the angular momentum number $l$ when the parameters $q_{i}$ are held constant. A similar trend is seen in Tables \ref{t2} and \ref{t4}, where the parameter $\alpha$ is fixed.
Regarding the imaginary part, its magnitude also increases as $l$ increases, for both scalar and electromagnetic perturbations, when either $q_{i}$ or $\alpha$ are kept constant. This indicates a shorter relaxation time, as the perturbations decay more rapidly. Since the imaginary part remains negative, the black holes are dynamically stable and do not develop growing modes.

\section{Absorption cross section}
\label{acs}

Another essential aspect of black holes is their accretion rate, which characterizes the process of absorbing surrounding matter and fields. Accretion plays a fundamental role in the astrophysical behaviour of black holes, particularly in the context of galactic nuclei.

In this section, we obtain the absorption cross section numerically as a sum of partial waves of a massless scalar wave. Considering the properties of the effective scattering potential \eqref{ec.ps}, we can notice that when $r_*\to -\infty$ and $r_*\to \infty$, the radial equation \eqref{ec.ts} takes the following form 
\begin{equation}\label{ecosaa}
\left(\frac{d^2}{dr_*^2}+\omega^2\right) R(r_*)=0,\text{ for }r_*\to -\infty\,,
\end{equation}
and 
\begin{equation}\label{ecosab}
	\left(\frac{d^2}{dr_*^2}+\omega^2\right) R(r_*)=0,\text{ for }r_*\to +\infty\,.
\end{equation}
The solutions of harmonic oscillator equations \eqref{ecosaa} and \eqref{ecosab} are ingoing and outgoing waves. For the scattering problem, we are interested in a solution that represents an incoming wave from infinity, a reflected wave towards infinity, and a transmitted wave absorbed by the black hole. Such solution takes the following form 
\begin{equation}\label{eqbcfd}
R(r_*)=\begin{cases}
A_{\text{in}}e^{-i\omega r_*}+A_{\text{out}}e^{i\omega r_*}\,,&\text{for }r_*\to+\infty\,,\\
e^{-i\omega r_*}\,,&\text{for }r_*\to-\infty\,,
\end{cases}
\end{equation}
where the $A_{\text{in}}$ represents the amplitude of an incoming wave at infinity and $A_{\text{out}}$ represents the amplitude of an outgoing wave at infinity. Such coefficients are related to the transmission amplitude and reflection amplitude as \begin{equation}
R_{\omega l}=\frac{A_{\text{out}}}{A_{\text{in}}}\,,\quad T_{\omega l}=\frac{1}{A_{\text{in}}}\,.
\end{equation} 
By using flux conservation, one can show that $R_{\omega l}$ and $T_{\omega l}$ satisfy the next relation
\begin{equation}
\left|R_{\omega l}\right|^2+\left|T_{\omega l}\right|^2=1\,.
\end{equation}
From the quantum mechanics, the total absorption cross section can be calculated as:
\begin{equation}\label{ec.sata}
\sigma_{\text{abs}}=\sum_{l=0}^{\infty}\sigma^{(l)}_{\text{abs}}\,,
\end{equation}
with $\sigma^{(l)}_{\text{abs}}$ the partial absorption cross section 
\begin{equation}
\sigma_{\text{abs}}^{(l)}=\frac{\pi}{\omega^2}\left(2l+1\right)\left|T_{\omega l}\right|^2\,.
\end{equation}
To obtain absorption cross section of scalar wave from the black hole, the first step is to solve \eqref{ec.ts} numerically from near the event horizon $r_+$ to a sufficiently large radius $r_0$ under the boundary conditions \eqref{eqbcfd}. Using the far-away behaviour given above, we determine $A_{\text{in}}$ and $A_{\text{out}}$ from $R_l(r_*)$ and $\frac{dR(r_*)}{dr_*}$. Thus, the reflection coefficient is then calculated.

\begin{figure}[!h]
	\centering
	\includegraphics[scale=0.93]{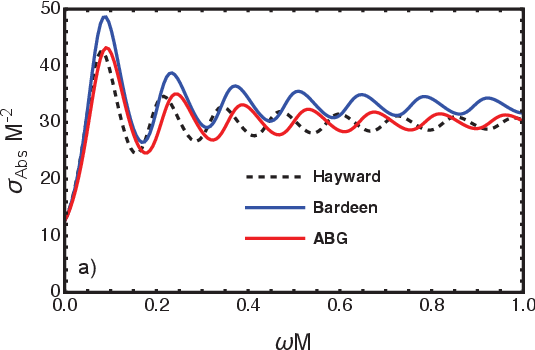}
	\includegraphics[scale=0.93]{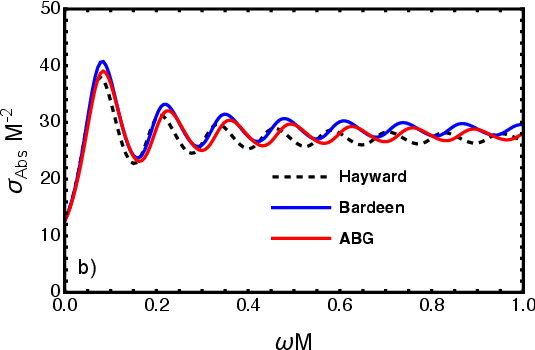}
	\caption{
	The behavior of total absorption cross section for a massless scalar field propagating in regular black hole surrounded by perfect fluid dark matter, with $\alpha=-0.2$. The summation in \eqref{ec.sata} is performed up to $l=10$. In the left panel we use $q_1^2=q_2^2=q_3^2=0.2$, while in the right panel $q_1^2=q_2^2=q_3^2=0.1$.}
	\label{f4}
\end{figure}

 Fig. \ref{f4} shows that the amplitudes of the absorption cross sections asymptotically approach a constant value as $\omega M$ increases. For fixed values of the parameter $q_{i}$, the Bardeen solution yields higher absorption cross-sections compared to the Hayward and ABG counterparts. Additionally, the difference among the curves becomes increasingly pronounced with larger $q_{i}$ values, indicating a significant dependence of the amplitude on $q_{i}$.

\begin{figure}[!h]
	\centering
	\includegraphics[scale=0.93]{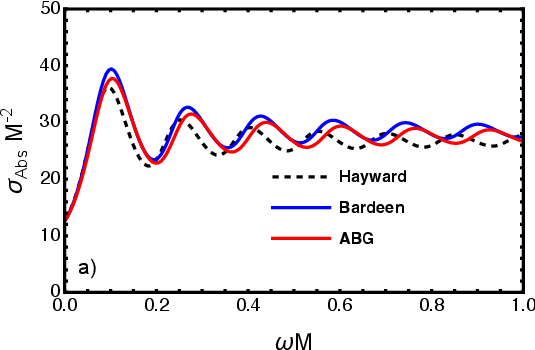}
	\includegraphics[scale=0.93]{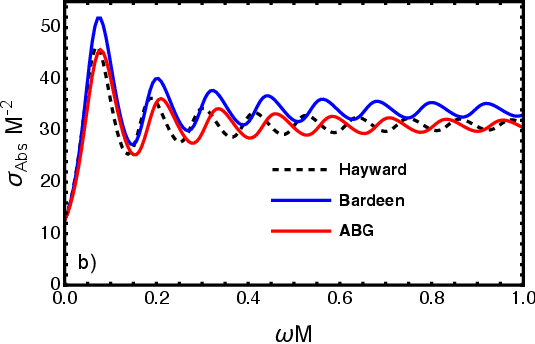}
	\caption{
	The behavior of total absorption cross section for a massless scalar field propagating in regular black hole surrounded by perfect fluid dark matter, with $q_{i}=0.15$. The summation in \eqref{ec.sata} is performed up to $l=10$. In the left panel we use $\alpha= -0.1$, while in the right panel $\alpha=-0.3$.}
	\label{f5}
\end{figure}

Finally in the Fig. \ref{f5}, we consider $\sigma_{abs}$ of the different BHs, as a function of the parameter $\alpha$,  and is possible to mention that for each BHs, $\sigma_{abs_{-0.3}}>\sigma_{abs_{-0.1}}$. Thus, the parameter $\alpha$ attenuates the absorption effect.

\section{Conclusions}\label{conclu}

The scalar, electromagnetic and gravitational perturbations were studied in three families of regular black holes: Hayward, Bardeen, and ABG immersed in PFDM. By employing both the sixth-order WKB method and the AIM, we determined the corresponding quasi-normal modes  and absorption cross sections, thus complementing previous analyses of regular black holes immersed in dark energy and dark matter.

By fixing the horizon radius at $r_{+}=1$, we analyze the dependence on the parameters ($q_{1},q_{2},q_{3}$) and on the PFDM intensity parameter, determining the ranges for which the black holes considered admit one or two horizons. The results indicate that $\alpha$ directly influences the effective potentials: more negative values of $\alpha$ lower the height of the potential barriers, thereby modifying the absorption cross section. This trend is consistent with earlier studies of black holes surrounded by quintessence and confirms that external matter distributions leave observable imprints.

The quasi-normal modes obtained exhibit negative imaginary parts in all cases, confirming that the black holes remain dynamically stable under scalar, electromagnetic and gravitational perturbations. Both the real and imaginary parts of the frequencies increase with the multipole number $l$, corresponding to higher oscillation frequencies and faster damping times. This behaviour, also reported in other works on regular  black holes, supports the generality of these features across different black hole models.

A comparative assessment of WKB and AIM reveals that while both methods converge for large $l$, WKB becomes less reliable in the fundamental modes. AIM, in contrast, requires $35$ iterations to achieve convergence, offers shorter computational times and produces stable results across a wide range of parameters. 

In the analysis of absorption cross sections, the Bardeen solution consistently exhibits higher absorption cross sections than the Hayward and ABG cases under comparable conditions. This suggests that different regular black holes may produce distinct scattering signatures. Then, such differences could, in principle, affect the energy dissipation rates of surrounding fields. These effects might eventually be probed indirectly through precise gravitational wave observations or black hole shadow measurements.

In summary, the results demonstrate that the presence of PFDM significantly modifies the quasi-normal modes (dynamical properties) and absorption cross section of regular black holes, while maintaining their stability. This strengthens the view that dark matter environments can leave measurable signatures in strong gravity regimes. This type of analysis seeks to connect theoretical predictions more directly with the observational frontier of black hole astrophysics.

\section*{ACKNOWLEDGMENT}

The authors acknowledge SNII--SECIHTI, M\'exico. 

\bibliographystyle{unsrt}
\bibliography{bibliografia}

@article{Planck:2018vyg,
    author = "Aghanim, N. and others",
    collaboration = "Planck",
    title = "{Planck 2018 results. VI. Cosmological parameters}",
    eprint = "1807.06209",
    archivePrefix = "arXiv",
    primaryClass = "astro-ph.CO",
    doi = "10.1051/0004-6361/201833910",
    journal = "Astron. Astrophys.",
    volume = "641",
    pages = "A6",
    year = "2020",
    note = "[Erratum: Astron.Astrophys. 652, C4 (2021)]"
}

@article{Planck:2019nip,
    author = "Aghanim, N. and others",
    collaboration = "Planck",
    title = "{Planck 2018 results. V. CMB power spectra and likelihoods}",
    eprint = "1907.12875",
    archivePrefix = "arXiv",
    primaryClass = "astro-ph.CO",
    doi = "10.1051/0004-6361/201936386",
    journal = "Astron. Astrophys.",
    volume = "641",
    pages = "A5",
    year = "2020"
}

@article{Kiselev:2002dx,
    author = "Kiselev, V.V.",
    title = "{Quintessence and black holes}",
    eprint = "gr-qc/0210040",
    archivePrefix = "arXiv",
    doi = "10.1088/0264-9381/20/6/310",
    journal = "Class. Quant. Grav.",
    volume = "20",
    pages = "1187--1198",
    year = "2003"
}

@article{Ghaderi:2017wvl,
    author = "Ghaderi, K.",
    title = "{Geodesics of black holes with dark energy}",
    doi = "10.1007/s10509-017-3204-y",
    journal = "Astrophys. Space Sci.",
    volume = "362",
    number = "12",
    pages = "218",
    year = "2017"
}

@article{Pedraza:2020uuy,
author = "Pedraza, Omar and L\'opez, L. A. and Arceo, R. and Cabrera-Munguia, I.",
title = "{Geodesics of Hayward black hole surrounded by quintessence}",
eprint = "2008.00061",
archivePrefix = "arXiv",
primaryClass = "gr-qc",
doi = "10.1007/s10714-021-02798-z",
journal = "Gen. Rel. Grav.",
volume = "53",
number = "3",
pages = "24",
year = "2021"
}

@article{Saleh:2018hba,
    author = "Saleh, Mahamat and Thomas, Bouetou Bouetou and Kofane, Timoleon Crepin",
    title = "{Quasinormal modes of gravitational perturbation around regular Bardeen black hole surrounded by quintessence}",
    doi = "10.1140/epjc/s10052-018-5818-9",
    journal = "Eur. Phys. J. C",
    volume = "78",
    number = "4",
    pages = "325",
    year = "2018"
}

@article{Kiselev:2003ah,
    author = "Kiselev, V. V.",
    title = "{Quintessential solution of dark matter rotation curves and its simulation by extra dimensions}",
    eprint = "gr-qc/0303031",
    archivePrefix = "arXiv",
    month = "3",
    year = "2003"
}

@article{Li:2012zx,
    author = "Li, Ming-Hsun and Yang, Kwei-Chou",
    title = "{Galactic Dark Matter in the Phantom Field}",
    eprint = "1204.3178",
    archivePrefix = "arXiv",
    primaryClass = "astro-ph.CO",
    reportNumber = "CYCU-HEP-12-04",
    doi = "10.1103/PhysRevD.86.123015",
    journal = "Phys. Rev. D",
    volume = "86",
    pages = "123015",
    year = "2012"
}

@article{Zhang:2020mxi,
    author = "Zhang, He-Xu and Chen, Yuan and Ma, Tian-Chi and He, Peng-Zhang and Deng, Jian-Bo",
    title = "{Bardeen black hole surrounded by perfect fluid dark matter}",
    eprint = "2007.09408",
    archivePrefix = "arXiv",
    primaryClass = "gr-qc",
    doi = "10.1088/1674-1137/abe84c",
    journal = "Chin. Phys. C",
    volume = "45",
    number = "5",
    pages = "055103",
    year = "2021"
}

@article{Das:2021otl,
    author = "Das, Anish and Saha, Ashis and Gangopadhyay, Sunandan",
    title = "{Study of circular geodesics and shadow of rotating charged black hole surrounded by perfect fluid dark matter immersed in plasma}",
    eprint = "2110.11704",
    archivePrefix = "arXiv",
    primaryClass = "gr-qc",
    doi = "10.1088/1361-6382/ac50ed",
    journal = "Class. Quant. Grav.",
    volume = "39",
    number = "7",
    pages = "075005",
    year = "2022"
}

@article{Abbas:2023pug,
    author = "Abbas, G. and Ali, R. H.",
    title = "{Thermal fluctuations, quasi-normal modes and phase transition of the charged AdS black hole with perfect fluid dark matter}",
    eprint = "2305.05541",
    archivePrefix = "arXiv",
    primaryClass = "gr-qc",
    doi = "10.1140/epjc/s10052-023-11580-1",
    journal = "Eur. Phys. J. C",
    volume = "83",
    number = "5",
    pages = "407",
    year = "2023"
}

@article{Anjum:2023axh,
    author = "Anjum, Arshia and Afrin, Misba and Ghosh, Sushant G.",
    title = "{Investigating effects of dark matter on photon orbits and black hole shadows}",
    eprint = "2301.06373",
    archivePrefix = "arXiv",
    primaryClass = "gr-qc",
    doi = "10.1016/j.dark.2023.101195",
    journal = "Phys. Dark Univ.",
    volume = "40",
    pages = "101195",
    year = "2023"
}

@article{Konoplya:2025mvj,
    author = "Konoplya, Roman A. and Khrabustovskyi, Andrii and K{\v{r}}{\'\i}{\v{z}}, Jan and Zhidenko, Alexander",
    title = "{Quasinormal ringing and shadows of black holes and wormholes in dark matter-inspired Weyl gravity}",
    eprint = "2501.16134",
    archivePrefix = "arXiv",
    primaryClass = "gr-qc",
    doi = "10.1088/1475-7516/2025/04/062",
    journal = "JCAP",
    volume = "04",
    pages = "062",
    year = "2025"
}

@article{Konoplya:2022hbl,
    author = "Konoplya, R. A. and Zhidenko, A.",
    title = "{Solutions of the Einstein Equations for a Black Hole Surrounded by a Galactic Halo}",
    eprint = "2202.02205",
    archivePrefix = "arXiv",
    primaryClass = "gr-qc",
    doi = "10.3847/1538-4357/ac76bc",
    journal = "Astrophys. J.",
    volume = "933",
    number = "2",
    pages = "166",
    year = "2022"
}

@article{Pedraza:2021hzw,
    author = "Pedraza, Omar and L\'opez, L. A. and Arceo, R. and Cabrera-Munguia, I.",
    title = "{Quasinormal modes of the Hayward black hole surrounded by quintessence: Scalar, electromagnetic and gravitational perturbations}",
    eprint = "2111.06488",
    archivePrefix = "arXiv",
    primaryClass = "gr-qc",
    doi = "10.1142/S0217732322500572",
    journal = "Mod. Phys. Lett. A",
    volume = "37",
    number = "09",
    pages = "2250057",
    year = "2022"
}

@article{Jusufi:2019ltj,
    author = "Jusufi, Kimet",
    title = "{Quasinormal Modes of Black Holes Surrounded by Dark Matter and Their Connection with the Shadow Radius}",
    eprint = "1912.13320",
    archivePrefix = "arXiv",
    primaryClass = "gr-qc",
    doi = "10.1103/PhysRevD.101.084055",
    journal = "Phys. Rev. D",
    volume = "101",
    number = "8",
    pages = "084055",
    year = "2020"
}

@article{Xu:2017bpz,
    author = "Xu, Zhaoyi and Wang, Jiancheng and Hou, Xian",
    title = "{Kerr\textendash{}anti-de Sitter/de Sitter black hole in perfect fluid dark matter background}",
    eprint = "1711.04538",
    archivePrefix = "arXiv",
    primaryClass = "gr-qc",
    doi = "10.1088/1361-6382/aabcb6",
    journal = "Class. Quant. Grav.",
    volume = "35",
    number = "11",
    pages = "115003",
    year = "2018"
}

@article{Lopez:2021ujg,
    author = "L\'opez, L. A. and Pedraza, Omar",
    title = "{Effects of quintessence on scattering and absorption sections of black holes}",
    eprint = "2103.06411",
    archivePrefix = "arXiv",
    primaryClass = "gr-qc",
    doi = "10.1007/s12648-022-02373-5",
    journal = "Indian J. Phys.",
    volume = "97",
    number = "1",
    pages = "285--294",
    year = "2023"
}

@article{Ramirez:2021ibk,
    author = "Ram\'\i{}rez, Valeria and L\'opez, L. A. and Pedraza, Omar and Ceron, V. E.",
    title = "{Scattering and absorption cross sections of Schwarzschild\textendash{}anti-de Sitter black hole with quintessence}",
    eprint = "2107.01282",
    archivePrefix = "arXiv",
    primaryClass = "gr-qc",
    doi = "10.1139/cjp-2021-0269",
    journal = "Can. J. Phys.",
    volume = "100",
    number = "2",
    pages = "112--118",
    year = "2022"
}

@article{dePaula:2023muc,
    author = "de Paula, Marco A. A. and Leite, Luiz C. S. and Crispino, Lu{\'\i}s C. B.",
    title = "{Geodesic analysis, absorption and scattering in the static Hayward spacetime}",
    eprint = "2311.15771",
    archivePrefix = "arXiv",
    primaryClass = "gr-qc",
    month = "11",
    year = "2023"
}

@article{Rincon:2025buq,
    author = "Rinc{\'o}n, {\'A}ngel and Fernando, Sharmanthie and Panotopoulos, Grigoris and Balart, Leonardo",
    title = "{Quasinormal modes and absorption cross-section of a Bardeen black hole surrounded by perfect fluid dark matter in four dimensions}",
    eprint = "2504.05215",
    archivePrefix = "arXiv",
    primaryClass = "gr-qc",
    month = "4",
    year = "2025"
}

@article{Paula:2020yfr,
    author = "Paula, Marco A. A. and Leite, Luiz C. S. and Crispino, Lu{\'\i}s C. B.",
    title = "{Electrically charged black holes in linear and nonlinear electrodynamics: Geodesic analysis and scalar absorption}",
    eprint = "2011.08633",
    archivePrefix = "arXiv",
    primaryClass = "gr-qc",
    doi = "10.1103/PhysRevD.102.104033",
    journal = "Phys. Rev. D",
    volume = "102",
    number = "10",
    pages = "104033",
    year = "2020"
}

@article{Hayward:2005gi,
      author         = "Hayward, Sean A.",
      title          = "{Formation and evaporation of regular black holes}",
      journal        = "Phys. Rev. Lett.",
      volume         = "96",
      year           = "2006",
      pages          = "031103",
      doi            = "10.1103/PhysRevLett.96.031103",
      eprint         = "gr-qc/0506126",
      archivePrefix  = "arXiv",
      primaryClass   = "gr-qc",
      SLACcitation   = "%%CITATION = GR-QC/0506126;%%"
}

@article{Bardeen1,
      author         = "J. M. Bardeen.",
      journal        = "in Conference Proceedings of GR5 (Tbilisi, USSR)",
      year           = "1968",
      pages          = "174"
}

@article{Ayon-Beato:1998hmi,
    author = "Ayon-Beato, Eloy and Garcia, Alberto",
    title = "{Regular black hole in general relativity coupled to nonlinear electrodynamics}",
    eprint = "gr-qc/9911046",
    archivePrefix = "arXiv",
    doi = "10.1103/PhysRevLett.80.5056",
    journal = "Phys. Rev. Lett.",
    volume = "80",
    pages = "5056--5059",
    year = "1998"
}

@article{Liu:2021fzr,
    author = "Liu, Peng and Niu, Chao and Zhang, Cheng-Yong",
    title = "{Linear instability of charged massless scalar perturbation in regularized 4D charged Einstein-Gauss-Bonnet anti de-Sitter black holes}",
    doi = "10.1088/1674-1137/abd01d",
    journal = "Chin. Phys. C",
    volume = "45",
    number = "2",
    pages = "025111",
    year = "2021"
}

@article{Liu:2020evp,
    author = "Liu, Peng and Niu, Chao and Zhang, Cheng-Yong",
    title = "{Instability of regularized 4D charged Einstein-Gauss-Bonnet de-Sitter black holes}",
    eprint = "2004.10620",
    archivePrefix = "arXiv",
    primaryClass = "gr-qc",
    doi = "10.1088/1674-1137/abcd2d",
    journal = "Chin. Phys. C",
    volume = "45",
    number = "2",
    pages = "025104",
    year = "2021"
}

@article{Konoplya:2003ii,
    author = "Konoplya, R. A.",
    title = "{Quasinormal behavior of the d-dimensional Schwarzschild black hole and higher order WKB approach}",
    eprint = "gr-qc/0303052",
    archivePrefix = "arXiv",
    doi = "10.1103/PhysRevD.68.024018",
    journal = "Phys. Rev. D",
    volume = "68",
    pages = "024018",
    year = "2003"
}

@article{Konoplya:2019hlu,
    author = "Konoplya, R. A. and Zhidenko, A. and Zinhailo, A. F.",
    title = "{Higher order WKB formula for quasinormal modes and grey-body factors: recipes for quick and accurate calculations}",
    eprint = "1904.10333",
    archivePrefix = "arXiv",
    primaryClass = "gr-qc",
    doi = "10.1088/1361-6382/ab2e25",
    journal = "Class. Quant. Grav.",
    volume = "36",
    pages = "155002",
    year = "2019"
}

@article{Konoplya:2011qq,
    author = "Konoplya, R. A. and Zhidenko, A.",
    title = "{Quasinormal modes of black holes: From astrophysics to string theory}",
    eprint = "1102.4014",
    archivePrefix = "arXiv",
    primaryClass = "gr-qc",
    doi = "10.1103/RevModPhys.83.793",
    journal = "Rev. Mod. Phys.",
    volume = "83",
    pages = "793--836",
    year = "2011"
}

@article{Nomura:2005dn,
    author = "Nomura, Hidefumi and Tamaki, Takashi",
    title = "{Continuous area spectrum in regular black hole}",
    eprint = "hep-th/0504059",
    archivePrefix = "arXiv",
    reportNumber = "WU-AP-219-05",
    doi = "10.1103/PhysRevD.71.124033",
    journal = "Phys. Rev. D",
    volume = "71",
    pages = "124033",
    year = "2005"
}

\end{document}